\documentclass{article}%
\usepackage{amssymb}
\usepackage{amsfonts}
\usepackage{amsmath}
\usepackage{graphicx}%
\providecommand{\U}[1]{\protect\rule{.1in}{.1in}}

\begin{document}

\title{Duane - Hunt Relation Improved}
\author{Ivan Cardenas and Anton Lipovka\\
\textit{Department of Research for Physics, Sonora University, }\\
\textit{83000, Hermosillo}, \textit{Sonora, M\'{e}xico}}
\maketitle

\begin{abstract}
In present paper the Duane-Hunt relation for direct measurement of the Planck
constant is improved by including relativistic corrections. New relation to
determine the Planck constant, suggested in this paper contains Duane-Hunt
relation as first term and can be applied in a wide range of energies.

\end{abstract}


 \smallskip\ 

\section{Introduction}

The origin of quantization and nature of the Planck constant remain the most
intriguing problems of physics, from the very moment of the creation of the
Quantum Theory (QT) at the beginning of the last century. From the beginning a
great number of attempts to explain the nature of the Planck constant were
made. One interesting way to do this was that suggested by Boyer in 1978
\cite{1} (see also \cite[-4]{2} and references therein), within the framework
of \textquotedblleft Zero-Point Radiation\textquotedblright. Another idea that
should be mentioned here was recently implemented in the framework of the
electrodynamics model of the atom proposed by M. Percovac \cite{5}. Within
this approach (as it was shown by author) the Planck constant value could
probably depends on the energy (see also experimental papers cited below in
which such a dependence was discussed).

In our papers \cite{6}, \cite{7}, \cite{8} we suggest a more natural manner to
explain the origin of the Planck constant due to the geometrical quantization
of action. This way not only allows us to calculate the correct value of the
Planck constant from the first principles (i.e. from cosmological parameters,
or from the geometry of our Universe), but it also allows us to unify the QT
and the relativistic physics. Moreover, from the observational fact that
geometry of our universe is changed adiabatically on time, it immediately
follows that the Planck constant should change its value with time. It is
important also to mention here the work of V. Garcia Morales \cite{9} in
whish, starting from the thermodynamics, he argued that neither time nor space
needs to be discrete but it is just action what is quantized. All this facts
suggest the importance of accurate measurement of Planck's constant in all
energy ranges for the possible detection of more subtle effects and, as a
consequence, for experimental justification of the theory.

Recently there appears a compilation of experimental measurements of the
Planck's constant for wide range of energies \cite[11-25]{10}, where the
probable experimentally measurable dependence of the Planck constant on energy
was supposed. So, more careful experiments over all energy ranges are clearly needed.

At the moment there are a lot of measurements of the Planck constant at small
energies from $0.001$ to $1eV$ \cite[11-18]{10}. Also the measurements for
$1MeV$ are available \cite[20-23]{19}, but in the range of energies from $1eV$
to $1MeV$ there are no precise data reported \cite[11-23]{10}. In this case it
is of great importance not only elaborate new experimental technique, but also
increase precision of actually available methods.

One such method of direct measurement of the Planck constant should be
mentioned here is the technique based on the Duane-Hunt (D-H) relation. This
relation was written from classical point of view, without relativistic
corrections. For this reason it shows dramatic discrepancy for the measured
Planck constant in respect to the CODATA value \cite[20-22]{19} which appears
in second - third digit. For this reason the D-H relation in its classical
form cannot be used for precise measurements of the Planck constant
\cite[25-27]{24}.

In present paper we have revised D-H relation by considering relativistic
corrections. An exact relativistic relation between the incident electron
energy and energy of produced X-ray photon is obtained, and 4 terms of its
expansion are suggested. This relation is precise if compared with the Duane
-- Hunt expression, which actually corresponds to the first term of the
expansion on parameter v/c.

\section{Bremsstrahlung}

Let's consider in details the geometry and physics of the process of
scattering of a relativistic electron on a tungsten atom with the emission of
a photon. In the Minkowski spacetime we represent it in fig.1.


\begin{center}
\begin{figure}[h]
\includegraphics[width=8.0cm,height=6.0cm]{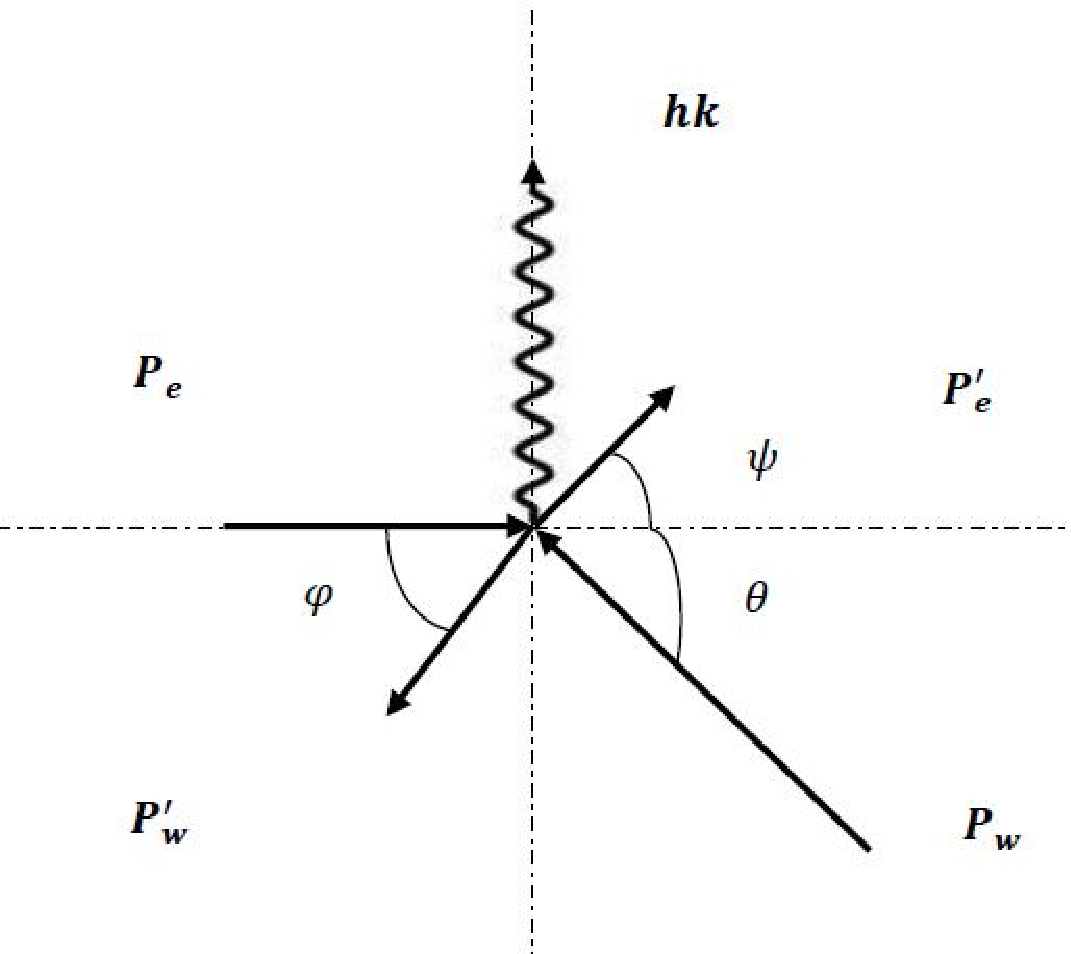}
\caption{Schematic
representation of the momentum four-vectors of each of the entities involved,
where the subscript $(e)$ represents the electron and the subscript $(w)$
corresponds to Tungsten, the prima superscript indicates the quantities after
collision, and $hk$ is the momentum of the emitted photon.}%
\label{Fig:fig_1}%
\end{figure}
\end{center}


In this case the conservation equations for temporal and spatial parts of the
$4$-momentum \cite{28} are%

\begin{equation}
E_{e}+E_{w}=E_{e}^{\prime}+E_{w}^{\prime}+E_{\gamma}\text{ \ \ ,} \tag{1}%
\end{equation}

\begin{equation}
\boldsymbol{p}_{e}+\boldsymbol{p}_{w}=\boldsymbol{p}_{e}^{\prime
}+\boldsymbol{p}_{w}^{\prime}+\boldsymbol{p}_{\gamma}\text{ \ \ ,} \tag{2}%
\end{equation}

were the subscript $\gamma$ denotes the emitted photon. We can rewrite $Eq(2)$
in the corresponding projections as follows:%

\begin{equation}
p_{e}-p_{w}\cos\theta=p_{e}^{\prime}\cos\psi-p_{w}^{\prime}\cos\varphi\text{
\ \ ,} \tag{3}%
\end{equation}

\begin{equation}
p_{w}\sin\theta=p_{e}^{\prime}\sin\psi-p_{w}^{\prime}\sin\varphi+\hbar k\text{
\ \ ,} \tag{4}%
\end{equation}
where $\hbar$ is the reduced Planck's constant, $k$ is the wavenumber of the
emitted photon, and $p$, $p^{\prime}$ are the $3$- moments of each constituent
of the system before $(p)$ and after $(p^{\prime})$ interaction.

In consequence with experimental evidence and D-H relation, we also assume
that the initial electron transfers all of its energy and momentum to the
photon (this is the case in the framework of a Duane -- Hunt relation). i.e.
$p_{e}^{\prime}\ll p_{e}$ and $p_{e}^{\prime}\approx0$. Thus $Eq.(3)$ and
$(4)$ take the form%

\begin{equation}
p_{e}-p_{w}\cos\theta=-p_{w}^{\prime}\cos\varphi\text{ \ \ ,} \tag{5}%
\end{equation}

\begin{equation}
p_{w}\sin\theta=\hbar k-p_{w}^{\prime}\sin\varphi\text{ \ \ .} \tag{6}%
\end{equation}
By squaring $Eq.(5)$ , we get%

\begin{equation}
p_{w}^{\prime2}\cos^{2}\varphi=\left(  p_{w}\cos\theta-p_{e}\right)
^{2}\text{ \ \ ,} \tag{7}%
\end{equation}
and rewrite $Eq.(6)$ to have%

\begin{equation}
\hbar k=p_{w}\sin\theta+p_{w}^{\prime}\sin\varphi\text{ \ \ .} \tag{8}%
\end{equation}
From $Eq.(7)$ through a trigonometric identity, we obtain%

\begin{equation}
p_{w}^{\prime2}\cos^{2}\varphi=p_{w}^{\prime2}-\left(  \hbar k-p_{w}\sin
\theta\right)  ^{2}\text{ \ \ ,} \tag{9}%
\end{equation}
and from $Eqs.(7)$ and $(9)$, we have%

\begin{equation}
\left(  p_{w}\cos\theta-p_{e}\right)  ^{2}=p_{w}^{\prime2}-\left(  \hbar
k-p_{w}\sin\theta\right)  ^{2}\text{ \ \ .} \tag{10}%
\end{equation}

Besides that, the relativistic equations for energy also should be applied.
They are $E_{e}^{2}=m^{2}c^{4}+p^{2}c^{2}$ and $E_{w}=K_{w}+mc^{2}$, where $E$
is the total energy, $K$ kinetic energy, and $p$ is the $3$-momentum. From
this relation and $Eq.(1)$ we have the following expression for energy:%

\begin{equation}
eU+m_{e}c^{2}+K_{w}+Mc^{2}=K_{e}^{\prime}+m_{e}c^{2}+K_{w}^{\prime}%
+Mc^{2}+h\nu\text{ \ \ ,} \tag{11}%
\end{equation}
where by assumption $V_{e}^{\prime}\ll V_{e}$ (it implies that $K_{e}^{\prime
}\approx0$). Therefore%

\begin{equation}
eU+K_{w}=K_{e}^{\prime}+K_{w}^{\prime}+h\nu\text{ \ \ .} \tag{12}%
\end{equation}

When the electron reaches the anode, it has a total energy $E_{e}$, whereas
$K_{w}$, $K_{w}^{\prime}$ are nonrelativistic, namely they can be substituted
by classical expression $\frac{p^{2}}{2m}$, then%

\begin{equation}
eU+\frac{p_{w}^{2}}{2M}=h\nu+\frac{p_{w}^{\prime2}}{2M}\text{ \ \ .} \tag{13}%
\end{equation}

Hence, the square of momentum of the Tungsten nucleus after the interaction is%

\begin{equation}
p_{w}^{\prime2}=2M\left(  eU+\frac{p_{w}^{2}}{2M}-h\nu\right)  \text{ \ \ .}
\tag{14}%
\end{equation}

The momentum of the electron $p_{e}$ in terms of its total energy is%

\begin{equation}
p_{e}c=\sqrt{\left(  eU\right)  ^{2}+2m_{e}c^{2}eU}\text{ \ \ .} \tag{15}%
\end{equation}

Substituting $Eqs.(14)$ and $(15)$ into $Eq.(10)$ and solving it for $h\nu$,
one can obtain the following expression,%

\begin{equation}
{\small h\nu=}\left(  Mc^{2}-p_{w}c\sin\theta\right)  \left[  -1+\sqrt
{1+\tfrac{2Mc^{2}eU+2p_{w}c\cos\theta\sqrt{\left(  eU\right)  ^{2}+2m_{e}%
c^{2}eU}-\left(  eU\right)  ^{2}-2m_{e}c^{2}eU}{\left(  Mc^{2}-p_{w}%
c\sin\theta\right)  ^{2}}}\right]  \tag{16}%
\end{equation}
\bigskip This expression was obtained in the approximation of the smallness of
the relativistic corrections to the classical energy of the tungsten atom.
Let's evaluate how accurate this expression is. For the electron, we used
exact relativistic expressions for energy and momentum, while for tungsten we
limited ourselves to the second term of the expansion (the kinetic energy of
the atom). The expansion for the energy of the tungsten atom up to the third
term is: $E_{w}=Mc^{2}+(Mc^{2}/2)(V/c)^{2}+(3/8)(Mc^{2}/2)(V/c)^{4}$. For
thermal velocities $V=10^{5}$ that correspond to a temperature of 2000 K, we
obtain the estimate for this correction: $\delta E/E=\delta h/h=(3/8)(V/c)^{4}%
=3\cdot10^{-23}$. As one can see, this is quite sufficient for accurate
measurements of the Planck constant.

Finally, to compare our result with the original D-H relation $h\nu=eU$\ , we
should expand this expression in respect to the small parameter $V/c$. By
leaving the first order terms in the expansion, we immediately obtain%

\begin{equation}
h\nu=eU+\frac{p_{w}\cos\theta\sqrt{\left(  eU\right)  ^{2}+2m_{e}c^{2}eU}}%
{Mc}-\frac{\left(  eU\right)  ^{2}}{2Mc^{2}}-\frac{m_{e}}{M}\left(  eU\right)
\text{ \ \ .} \tag{17}%
\end{equation}

So, as one can see the first term coincides exactly with the D-H relation,
while the subsequent three terms are first-order relativistic corrections to
the relation.

\section{Conclusion}

In present paper the improved expression for the D-H relation is obtained.
This expression takes into account the fact that velocities of atoms are
differ from zero and relativistic corrections for the electron energy-momentum
are significant. We suggest exact formula (16), which generalizes the D-H
relation by taking into account relativistic effects. We also suggest
expansion of the relation obtained, in order to compare our result with the
original D-H relation. As one can see, the first term of this expansion
coincide precisely with the Duane-Hunt law $h\nu=eU$, and the next terms
correspond to the relativistic corrections.

Obtained expression allows us to determine the Planck's constant (with
appropriate experimental data) with greater precision and in great range of
energies (from 1eV to 1MeV) which was unavailable for other experiments. For
this reason one can hope that obtained expression will be useful to close this
great gup in experimental measurements of the Planck constant.

To conclude it should be mentioned again that there exist a large number of
physical theories, particularly based on the geometric approach \cite{6},
\cite{7}, \cite{8} and \cite[31]{30}, which not only allow us to calculate the
Planck constant from the first principles, but also predict its variation.
Thus, the problem of accurate experimental determination of Planck constant is
of crucial importance, since such measurements will make it possible to
discriminate a large number of theories by choosing those that correspond to
the experiment.

Anton Lipovka: aal@cifus.uson.mx.\\ 
Ivan Cardenas: cardenas@posgrado.cifus.uson.mx

\section{Acknowledgments}

Author (Ivan Cardenas) is grateful for the support provided by CONACyT.

\end{document}